\documentclass{aa}
\usepackage{graphicx}
\usepackage{epsfig}
\usepackage{txfonts}
\begin{document}

   \title{Azimuthal dependence of the density distribution
   in outer galactic discs accreting intergalactic flows}

   \subtitle{}

   \author{M. L\'opez-Corredoira$^1$, J. Betancort-Rijo\inst{1,2}}

   \offprints{martinlc@iac.es}

\institute{
$^1$ Instituto de Astrof\'\i sica de Canarias, E-38200, La 
Laguna, Tenerife, Spain\\
$^2$ Departamento de Astrof\'\i sica, Universidad de La Laguna, Tenerife,
Spain}

   \date{Received xxxx; accepted xxxx}

% \abstract{}{}{}{}{}
% 5 {} token are mandatory

  \abstract
   {}
   {The amplitude and scaleheight of the Galactic gas disc density 
   are not axisymmetric against expectations in a self-gravity 
   axisymmetric disc. However,
   this lopsidedness can be explained in terms of intergalactic accretion flows,
   which produce non-axisymmetric pressure on the disc. This mechanism
   could be also responsible for the formation of a warp.}
   {We analytically derive the relationship between the disc density and
   the self-gravity and external pressure.}
   {The same scenario of accretion as we proposed years ago to
   explain the formation of the warp explains the azimuthal dependence of the 
   density and its scaleheight, with minimum/maximum in the positions 
   of maximum amplitude of the warp ($\phi \approx 95^\circ $ and 
   275$^\circ $), as expected from its pressure distribution.}
   {}

\titlerunning{Lopsidedness due to IGM flows}
\authorrunning{L\'opez-Corredoira \& Betancort-Rijo}   

\maketitle
%
%________________________________________________________________

\section{Introduction}

There are several proofs that the scaleheight of the gas disc
in our Galaxy depends both on the galactocentric distance 
(Narayan \& Jog 2002; Nakanishi \& Sofue 2003) and the azimuth
(Voskes 1999; Levine et al. 2006; Kalberla et al. 2007).
And there are explanations for the radial dependence of the 
scaleheight of galactic discs in spiral galaxies (flares) in terms of
self-gravitating discs (e.g., Narayan \& Jog 2002) or magnetic fields
(Battaner \& Florido 1995). 
However, it is not easy to understand the azimuthal dependence
in the same terms, except for an exotic proposal of the existence of
a ring of dark matter embedded in the galactic disc with a radius that
depends on the azimuth (Kalberla et al. 2007). 
We think we have a better solution that is less exotic, 
less ad hoc, and that agrees with other observed galactic 
features, too. 

In L\'opez-Corredoira et al. (2002, hereafter LBB), we proposed 
a mechanism to explain the formation of warps 
(S-shaped or U-shaped or a combination of both)
in spiral galaxies: the accretion of the intergalactic medium (IGM) 
onto the disc (see LBB, Fig. 6).
Indeed, up to now there have been no alternative explanations 
for the formation of U-warps.
No massive halo is necessary, or high values of magnetic fields, or
satellite companions are necessary, although the presence 
of these elements would not modify qualitatively the present conclusions. 
In a Milky-Way-like galaxy, the mean density of baryonic matter in 
the IGM needed to produce the observed warp 
is around $10^{-25}$ kg/m$^3$ when the 
infall velocity at a large distance is $\sim 100$ km/s (LBB).
These numbers were corroborated independently by S\'anchez-Salcedo (2006).
This hypothetical low-density net flow is a very reasonable physical assumption
that would explain why most spiral galaxies are warped.
There are also different types of observations successfully 
explained by the LBB hypothesis (L\'opez-Corredoira et al. 2008): 
1) accretion of $\sim 1$ M$_\odot$/yr of low metallicity gas onto 
the disc as expected from the chemical evolution of the Milky Way
(see LBB, \S 3.1);
2) frequency of warps and its amplitude
depending on environment; 3) lower frequency of U-warps over S-warps. 

In this paper, we claim that the same mechanism proposed by
LBB to explain the Galactic S-warp is able to explain another
observational fact: the azimuthal dependence
of the gas density and scaleheight in the outer disc.

\section{Density distribution in the vertical direction of a self-gravitating
disc}
\label{.grav}    

As has been known for a long time (Spitzer 1942; Narayan \& Jog
2002 and references therein), the distribution
of matter in the vertical direction of the disc can be derived from
applying the hydrostatic equilibrium,
\begin{equation}
\frac{\langle v_z^2\rangle}{\rho }\frac{\partial \rho }{\partial z}=a_z
\label{hydro}
,\end{equation}
and the Poisson equation,
\begin{equation}
\frac{\partial a_{z,grav}}{\partial z}=
-4\pi G\rho -\frac{1}{R}\frac{\partial (R\ a_{R,grav})}{\partial R}
-\frac{1}{R}\frac{\partial a_{\phi ,grav}}{\partial \phi }
\label{poisson}
,\end{equation}
where $\rho $ is the mass density; $\langle v_z^2\rangle$ the
variance of the vertical velocities along the vertical $z$ direction
in an isothermal disc; $a_R$, $a_\phi $, and $a_z$ are the three components
of the acceleration per unit mass in cylindrical coordinates ($R$, $\phi $, 
$z$; the vertical direction $z$ being defined as perpendicular to the
ring); and $a_{x,grav}$ is the part of the $x$-component due to the 
self-gravity of the disc. On the right-hand side of Eq. (\ref{poisson}),
the third term is zero for an axisymmetric disc, and 
the second term is neglected for a thin-disc approximation, which are
the normal assumptions in most analyses of the scaleheight. 
Here we do not neglect the second term since the
scaleheight is not small. Instead, we adopt a monopolar 
approximation for the radial acceleration,
\begin{equation}
a_{R,grav}\approx \frac{-GM_{gal}}{R^2}
\label{gravrad}
,\end{equation}
where $M_{gal}$ is the mass of the Galaxy within radius $R$, which
we consider constant for the outer disc.
For the azimuthal acceleration $a_{\phi,grav}$, there is some 
contribution from the warp, but we did check that it is negligible.
We integrated over all the rings
of the warped galaxy with a height $z_w(R,\phi )$, which we take
from Levine et al. (2006), and we found that the azimuthal accelerations
are of the order of $10^{-13}$ m/s$^2$, and the third
term of the right-hand side Eq. (\ref{poisson}) is more than 50 times lower 
than the second term for all values of $R>15$ kpc and $\phi $.
Therefore we can neglect this direct effect of the warp. The
effect of the bar is also negligible: azimuthal accelerations
lower than $5\times 10^{-14}$ m/s$^2$ with the potential of the
long bar of L\'opez-Corredoira (2007, Eq. 7).
Therefore, if one only considers the self-gravity acceleration, by doing the
derivative with respect to $z$ of Eq. (\ref{hydro}) and assuming that
the gravity is the only acceleration dependent on $z$,
with Eqs. (\ref{poisson}) and (\ref{gravrad}), we get

\begin{equation}
\rho ''-\frac{\rho '^2}{\rho }+K_1\rho ^2+K_2\rho =0 
\label{diffeq}
,\end{equation}
where $K_1=\frac{4\pi G}{\langle v_z^2\rangle}$,
$K_2=\frac{GM_{gal}}{R^3\langle v_z^2\rangle}$.
The prime (') stands for the derivative
with respect to $z$.

\subsection{Solutions of the differential equation \protect{(\ref{diffeq})}}

The differential equation (\ref{diffeq}) may be expressed as an integral.
If we set as a boundary condition that the maximum density is reached
at $z=z_0$ ($\rho (z_0)=A$, $\rho '(z_0)=0$), we obtain
\begin{equation}
z=z_0\pm \int _0^{\frac{A-\rho (z)}{A}}
\frac{dx}{(1-x)\sqrt{2K_1Ax-2K_2ln (1-x)}}
,\end{equation}
where $x=\frac{A-\rho }{A}$; 
which is a symmetrical distribution of $\rho (z)$ with respect to $z_0$.
The half-width-half-maximum (HWHM) is
\begin{equation}
HWHM=\int _0^{1/2}\frac{dx}{(1-x)\sqrt{2K_1Ax-2K_2ln (1-x)}}
\label{hwhm}
,\end{equation}
and the surface density is
\begin{equation}
\sigma=\int _{-\infty}^{\infty }dz\rho (z)=2A
\int _0^1\frac{dx}{\sqrt{2K_1Ax-2K_2ln (1-x)}}
\label{sigma}
.\end{equation}

For very thin discs ($\rho $ very high), the second term on the right-hand 
side of Eq. (\ref{poisson}) is negligible ($K_2\approx 0$). 
The solution to this differential equation
with $K_2=0$ is the classical squared hyperbolic secant solution 
(Spitzer 1942): $\rho (z)=A\ sech^2[a(z-z_0)]$, 
$a=\sqrt{A K_1/2}=K_1\sigma /4$.

Since $K_1$ and $K_2$ do not depend on $\phi $, it is clear that
the width of the disc, HWHM, is independent of $\phi $, too, and only depends
on the radius, so the self-gravity application does not explain
the azimuthal dependence of the scaleheight.

\section{With external pressure by a continuous accretion of IGM}
\label{.LBB}

We propose that the explanation for the variation in the 
gas disc thickness depending on $\phi $
is that the external pressure due to the accretion of IGM onto
the disc depends on $\phi $,
because the average flow falls down to the disc with an 
angle different to $\pi /2$ in general 
with respect to the plane (LBB).
The pressure exerted over the disc would be similar to a piston 
mechanism, only from one side of the disc (S\'anchez-Salcedo 2006, \S 4.6).
The action is perhaps ram-pressure, due to the friction of clouds 
against the interstellar medium (Sofue \& Wakamatsu 1993). 

The vertical acceleration due to this pressure, $a_{z,pres}$, 
will depend on $z$. There is a gradient of force in the vertical
direction due the higher absorption of linear momentum by the first
layers of the disc that collide with the accreted gas.
For a very low dense, disc, as 
is the case of the very outer disc, the absorption of momentum 
is not total, because part
of the gas can cross the disc completely and escape
from it. We suppose that
the external pressure is attenuated exponentially along the
z-axis $P_{z,ext}(z)=P_{z,ext}(\pm \infty )
e^{\pm C[\int _{\pm \infty }^z\rho (z)dz]}$,
where $C$ is a cross-section per unit mass characteristic of the interaction 
between the accreted flow and the gas disc along the vertical axis.
Here we do not take into account the effect of the galactic rotation
and the variations in the attenuation with different incident angles. 
The sign + or - depends on whether the flow comes 
from $z=\infty $ or $z=-\infty $. The force per unit
volume is $-\nabla P_z(z)$. Hence,
\begin{equation}
a_{z,pres}(R,\phi ,z)
=-(\pm )F_z(R,\phi )C\ e^{\pm C[\int _{\pm \infty }^z\rho (z)dz]}
\label{azz}
,\end{equation}
where $F_z\equiv P_{z,ext}(\pm \infty )$ is the vertical component of 
the external force per unit surface ($dS=RdRd\phi$) due to accretion.

We do the derivative of Eq. (\ref{azz}), taking the exponential
close to one in a low-density disc in the very outer disc 
[$\frac{1}{2}C\sigma $ small compared to one; indeed, in \S \ref{.comp} we 
will see that $\frac{1}{2}C\sigma <\sim 10^{-13}
[\rho _b ({\rm kg/m^3})]^{-1/2}$, which is smaller than one for the
expected values of $\rho _b$ ($\sim 10^{-25}$  kg/m$^3$) 
although not much smaller; but we take this as a rough approximation]:
\begin{equation}
a_{z,pres}'\approx -F_zC^2\rho 
\label{derazz}
.\end{equation}

Given a continuous inflow of particles with density 
at infinite distance $\rho _b$, velocity $v_0$, and angle with respect to the
plane $\theta _0$ with azimuth $\phi _0$ of the direction of the
inflow (so the flow comes from $-\theta _0$, $\phi _0+\pi $), 
and Galactic mass within $R$ of $M_{gal}$, the following results hold 
[from LBB, Eq. (45); applying Eqs. (28), (33), (34), (38) of the same 
paper]\footnote{We have found an erratum in LBB: Eq. (28)
should have opposite sign, i.e. $e_{0Q}=-\cos (\vec{v_0},\vec{r_Q})=-
\cos (\theta _0)\cos (\phi _0-\phi)$. However, numbers and orientation 
of the warp are correct as stated in LBB.}

\begin{equation}
F_z(R,\phi)dS=\frac{\rho _bv_0^2dS}{R^2|\sin \theta _0|}
\frac{x_1}{\left(1+\sin ^2(\phi _0-\phi)\left(\frac{1}{\sin ^2(\theta _0)}
-1\right)\right)}
\label{verforce}
\end{equation}
\[\times
\left[\frac{Rx_2}{2}
+\sqrt{\frac{R^2x_2^2}{4}
+\frac{RGM_{gal}x_3}{v_0^2}}\right]^2
\times
\left[\frac{x_2}{2}
+\frac{\frac{Rx_2^2}{4}
+\frac{GM_{gal}x_3}{2v_0^2}}
{\sqrt{\frac{R^2x_2^2}{4}
+\frac{RGM_{gal}x_3}{v_0^2}}}\right]
,\]\[
x_1(\phi )=\sqrt{1-\cos ^2\theta _0\sin ^2(\phi _0-\phi )}
\]\[
x_2(\phi )=\sqrt{1-\cos ^2\theta _0\cos ^2(\phi _0-\phi )}
\]\[
x_3(\phi )=1+\cos \theta _0\cos (\phi _0-\phi )
.\]

\begin{figure}
\begin{center}
\vspace{.2cm}
\mbox{\epsfig{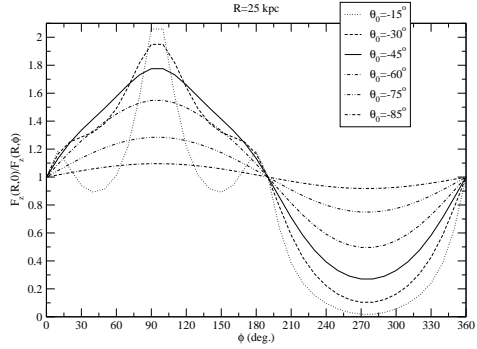}}
\end{center}
\caption{$F_z^{-1}$ according to eq. (\protect{\ref{verforce}}) with the
parameters given by LBB and $\phi _0=275^\circ$.}
\label{Fig:fz}
\end{figure}

Figure \ref{Fig:fz} plots $F_z^{-1}$ with
the parameters used by LBB: $M_{gal}=2\times 10^{11}$ M$_\odot $,
$v_0=100$ km/s, and $\theta _0$ is a free parameter.
$F_z$ is $\sim 10^{-14}$ kg/m/s$^2$.
We take $\phi _0=275^\circ $, $\theta _0<0$, the
corresponding direction of the accreted flow, in 
order to produce the Galactic S-warp with U-warp southwards (LBB)
and maximum of the S-warp at $\phi =95^\circ, 275^\circ $, as observed
(Voskes 1999).
The dependence is plotted for $R=25$ kpc and variable $\phi $,
while other values of $R$ give
different amplitudes in the variation, but similar azimuthal dependence.
The amplitude of the variation with the azimuth
is strongly dependent on $\theta _0$. Clearly, for any value 
of $\theta _0$, the minimum pressure
is for $\phi =\phi _0+\pi=95^\circ $, and the maximum is for $\phi =
\phi _0=275^\circ $. 
Nonetheless, if we took into account the rotation of the galaxy
and an external pressure attenuation dependent on this and the incident
angle, or the degree of clumpiness of IGM and disc, the difference 
in the position of the maximum and minimum would not be strictly $\pi $,
and the shape of Fig. \ref{Fig:fz} would vary.

\subsection{Differential equation}

If we now consider that the total acceleration is 
$a_z=a_{z,grav}+a_{z,pres}$ in Eq. (\ref{hydro}), assuming an 
immediate response of the disc to the pressure exerted by the accreted 
material (L\'opez-Corredoira et al. 2008), 
together with Eqs. (\ref{poisson}) and (\ref{derazz}),
we again get the same differential equation (\ref{diffeq}) but with
constants $K_1=\frac{4\pi G+F_z(R,\phi )C^2}{\langle v_z^2\rangle (R,\phi)}$,
$K_2=\frac{GM_{gal}}{R^3\langle v_z^2\rangle (R,\phi )}$.
The total acceleration is null at $z_0$, where the pressure acceleration
is compensated by other gravitational forces to keep the warp static.
With $C\sim 10^2-10^3$ m$^2$/kg, we get values of $F_z(R,\phi )C^2$ of 
the order of $4\pi G$, that is, an external pressure as significant
as the self-gravity. And, most important, HWHM will depend on $\phi $
because $F_z$ depends on $\phi $ and also on the maximum amplitude of the
density, $A$, and the dispersion of velocities depends on $\phi $ through
its dependence on the pressure at $z=z_0$ $P(\phi )$:
\begin{equation}
A(\phi)=A(\phi _0)\left(\frac{P(\phi)}{P(\phi _0)}\right)^{\frac{1}{\gamma}}
\label{amppres}
,\end{equation}\begin{equation}
\langle v_z^2\rangle (\phi)=
\langle v_z^2\rangle (\phi _0)\left(\frac{P(\phi)}
{P(\phi _0)}\right)^{\frac{\gamma -1}
{\gamma}}
\end{equation}\begin{equation}
P(\phi)\propto \left(1+\frac{F_z(\phi )C^2}{4\pi G}\right)  
.\end{equation}
This last proportionality of the pressure stems from 
the external pressure acting like an extra self-gravity, and
the pressure is proportional to the total acceleration.
For a monoatomic gas, $\gamma =5/3$
for adiabatic compression, and $\gamma =1$ for isothermal one. This applies
in the distribution of pressure as a function of $\phi $; for the vertical
dependence, we have already assumed that it is isothermal.

\begin{figure}
\begin{center}
\vspace{.2cm}
\mbox{\epsfig{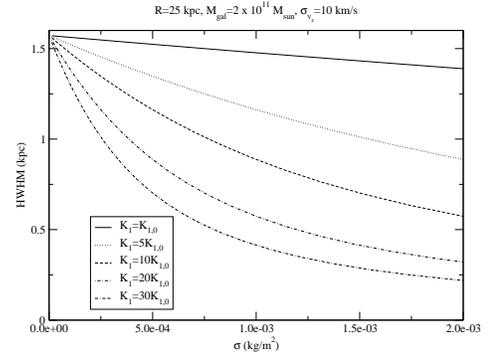}}
\end{center}
\caption{Dependence of HWHM on $\sigma $, from Eqs. (\protect{\ref{hwhm}}) 
and (\protect{\ref{sigma}}), for different values
of $K_1$ with the minimum value $K_{1,0}=\frac{4\pi G}{\langle v_z^2\rangle}$ 
given when only gravity is present and other values when an external 
pressure is added, and fixed $K_2$ and $\langle v_z^2\rangle$.}
\label{Fig:hwhm}
\end{figure}

In Fig. \ref{Fig:hwhm}, we see how the HWHM is reduced for a given
$\sigma $ when $K_1$ is increased, i.e. when $F_z$ is increased.
The parameters are $R=25$ kpc, $M_{gal}=2\times 10^{11}$ M$_\odot $,
and $\sigma _{v_z}=10$ km/s.

\section{Comparison with the observations}
\label{.comp}

Voskes (1999, Fig. 15) and Levine et al. (2006, Fig. 5)
have shown that the scaleheight of the outer disc ($R>20$ kpc) 
is $2-3$ times higher on average for $0<\phi <180^\circ$ than for 
$180^\circ <\phi <360^\circ $. A more accurate estimation of the
maximum of the scaleheight is derived by Kalberla et al. (2007, Figs. 18, 19), 
who place it for $90^\circ <\phi <110^\circ $, while the minimum for 
$250^\circ <\phi <270^\circ $. Within the LBB scenario,
two possible directions are possible to produce the observed S-warp,
and this is one of them: the wind coming from the direction
of the northern warp, although this solution could not explain the
asymmetry of the southern/northern warp as a sum of S+U warp.  
A higher pressure is expected for the region
around the southern warp and consequently a lower thickness therein.
Not only is the pressure lower at $\phi \approx 
90^\circ $, but the surface density $\sigma $ is also lower than for
the average value of $\sigma (R)$ (Voskes 1999, Fig. 13; 
Levine et al. 2006, Fig. 1; Kalberla \& Dedes 2008, Fig. 9)
and the amplitude of the density $A$ (Kalberla \& Dedes 2008, Fig. 8).
This is another fact explained by our model by means
of Eqs. (\ref{sigma}) and (\ref{amppres}): the lower the pressure the
lower the density. 

\begin{figure}
\begin{center}
\vspace{1cm}
\mbox{\epsfig{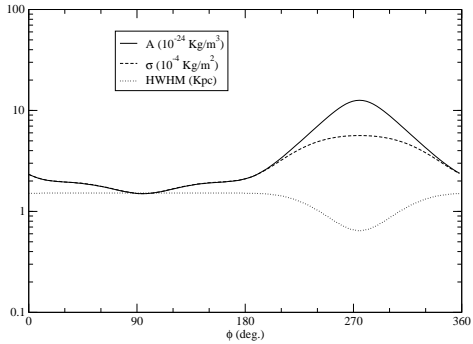}}
\end{center}
\caption{Azimuthal dependence of the variables HWHM,
$A$ and $\sigma $ for $R=25$ kpc. Parameters: $v_0=100$ km/s, $M_{gal}=2\times
10^{11}$ M$_\odot $, $\sigma _{v_z}(\phi _0)=10$ km/s, 
$A(\phi _0)= 1.5\times 10^{-24}$ kg/m$^3$, $\phi _0=275^\circ$;
$\gamma =1.22$, $\theta _0=-31^\circ $, 
$\rho _bC^2=1.8\times 10^{-19}$ m/kg.}
\label{Fig:azdep}
\end{figure}

If we take the values observed by Kalberla et al. (2007, 2008) at
$R=25$ Kpc, excluding the bin of $90^\circ <\phi <110^\circ $,
which is possibly an outlier in a special region: 
$A(\phi _0)= 1.5\times 10^{-24}$ kg/m$^3$; $\frac{A(\phi _0+\pi)}
{A(\phi _0)}\approx 7$, $\frac{\sigma(\phi _0+\pi)}{\sigma(\phi _0)}\approx 
4$, $\frac{HWHM(\phi _0)}{HWHM(\phi _0+\pi)}=2.5$. This last number is not
independent of the other three. The parameters
that better fit these numbers are: $\gamma =1.22$, $\theta _0=-31^\circ $,
$\rho _bC^2=1.8\times 10^{-19}$ m/kg, giving an
azimuthal dependence of HWHM, $A$ and $\sigma $ as plotted in
Fig. \ref{Fig:azdep}. The average accretion at $R$=25 kpc is 
$5\times 10^{-4}\left(\frac{\rho _b}{10^{-25}\ kg/m^3}\right)$ 
kg/m$^2$/Gyr, which is around 2 times the average surface density per Gyr; 
note, however, that only a small ratio of its linear momentum ($\sim C\sigma $) 
is transmitted to the Galactic disc.

For the shape of the scaleheight as a function of $\phi $ for $R=25$ kpc
(or any other value of the galactocentric distance $20\le R\le 40$ kpc),
we observe in Fig. 19 of Kalberla et al. (2007) a strongest dependence
of the scaleheight with $\phi $ near the maximum; 
however, the variation in amplitude is very slight around the 
minimum of flaring. This is not observed in the shape of Fig. 
\ref{Fig:azdep}. 

Therefore, at present with our simple model, we
cannot explain the exact azimuthal dependence. Possibly, our rough
assumptions in the present and LBB calculations need to be
improved to get a more accurate result, assuming that Kalberla
et al. (2007, 2008) are also correct with their assumptions on
the kinematics of the Galaxy. Indeed, the Kalberla et al. (2007, 2008)
density distributions are derived by assuming a non-axisymmetric disc 
using epicyclic corrections in the rotation curve, consistent with
their hypothesis to explain the asymmetries in the outer disc.
Since we can explain the asymmetries of the outer disc with
the accretion of IGM flows, it is possible that those 
epicyclic corrections are not necessary, although they are not
necessarily inconsistent with our scenario. 
In any case, our model describes the asymmetries roughly. 
The value of the parameters for the infall material also
more or less agree the values necessary for the warp
production in LBB. If Kalberla et al. assumed axisymmetric
rotation, they would get a gas distribution that is several kpc 
more extended to the south ($180^\circ <l<360^\circ $) 
than to the north (Levine et al. 2006, Fig. A18).
This would also agree with our scenario because 
we predict a higher density in the south in general.

\section{Conclusions}

The same LBB scenario that explains the formation of warps 
is able to directly explain the azimuthal dependence of the 
HI density distribution in the Milky Way,
with maximum and minimum in the positions expected,
without introducing ad hoc elements.
S\'anchez-Salcedo (2006) raises the criticism that LBB mechanism is not 
plausible because it would produce a dependence on the scaleheight of the disc 
with the Galactocentric azimuth, $\phi $, in the outer disc. Rather than being 
an objection, it is another argument in favour of our model
because this dependence is actually observed in our Galaxy.

\

{\bf Acknowledgments:}
Thanks are given to the anonymous referee for helpful comments, 
and to Joly Adams (language editor of A\&A) for proof-reading
this paper. MLC was supported by the {\it Ram\'on y Cajal} Programme
and the grant AYA2007-67625-CO2-01 of the Spanish Science Ministery.

\end{document}